\documentclass[final]{appolb}
\usepackage{amssymb,amsmath}
\usepackage{epsfig}

\begin{document}
\headauthor{Marcin Raczkowski}
\headtitle{$d$-wave Superconductivity, Orbital Magnetism,
and Unidirectional \ldots}
\title{$d$-wave Superconductivity, Orbital Magnetism, and Unidirectional 
Charge Order in the $t$-$J$ Model\thanks{Presented at the XIII National 
School ''Superconductivity, \ldots '', L\c{a}dek Zdr\'oj 2007}}
\author{Marcin Raczkowski
\address{Marian Smoluchowski Institute of Physics, Jagellonian
 University, \\ Reymonta 4, PL-30059 Krak\'ow, Poland}}
\maketitle

\begin{abstract}
Recent scanning tunneling microscopy in the superconducting regime of two 
different cuprate families has revealed unidirectional bond-centered
modulation in the local electronic density of states. Motivated by this 
result we investigate the emergence of modulated  $d$-wave superconductivity 
coexisting with charge domains that form along one of the crystal axes. 
While detailed stripe profiles depend on the used form of the Gutzwiller 
factors, the tendency towards a valence bond crystal remain robust. 
We also find closely related stripe phase originating from the staggered flux 
phase, a candidate for the pseudogap phase of lightly doped cuprates.

\end{abstract}
  
\section{\label{intro} Introduction}

It is now well established that the simplest model proposed to describe the 
physics of the high-$T_c$ superconductors, the so-called $t$-$J$ 
model~\cite{tJ}, 
\begin{equation}
{\cal H}= - t \sum_{\langle ij\rangle,\sigma}
     ({\tilde c}^{\dag}_{i\sigma}{\tilde c}^{}_{j\sigma} + h.c.)
      + J\sum_{\langle ij\rangle} {\bf S}_i \cdot {\bf S}_j,
\label{eq:tJ}
\end{equation}
where ${\tilde c}^{\dag}_{i\sigma}=(1-n_{i,-\sigma})c^{\dag}_{i\sigma}$ 
is the Gutzwiller projected electron operator and 
$n_{i\sigma}$ is the particle number operator, yields apart from the true 
long-range magnetic order, characteristic of the undoped parent Mott 
insulators, an array of quantum SU(2)-invariant ground states \cite{Sach03}. 
In fact, it is quite natural to expect that strong quantum fluctuations 
arising from both low spin $S=1/2$ of the copper ions and the two-dimensional 
nature of the CuO$_2$ planes, should lead to {\it quantum disordered} 
states with only short-range antiferromagnetic (AF) spin correlations. 
The most famous example of such states is a resonating valence bond (RVB) 
phase~\cite{And87}.
Remarkably, Anderson's RVB theory based on a Gutzwiller projected BCS trial wave
function, which parameters are usually determined either by using renormalized
mean field theory (RMFT)~\cite{RVB} or by Variational Monte Carlo (VMC) method
\cite{VMC}, not only predicted correctly the $d$-wave symmetry of the
superconducting (SC) order parameter~\cite{dwave}, but in addition,
it reproduced experimental doping dependence of a variety
of physical observables in the SC regime~\cite{th_exp}.

Moreover, the tendency towards valence bond amplitude maximization might 
enhance charge and spin stripe correlations in the $d$-wave RVB state. 
The presence of charge and long-range spin stripe order has been detected
in neutron scaterring experiments and confirmed in resonance $x$-ray
scattering in a few special cuprate compounds,
namely La$_{1.6-x}$Nd$_{0.4}$Sr$_x$CuO$_4$ and 
La$_{2-x}$Ba$_x$CuO$_4$~\cite{xray}. However, such stripe order competes 
with superconductivity and thus strongly reduces $T_c$ \cite{Valla}. 
In contrast, recent scanning tunneling microscopy (STM) on different cuprate 
families Ca$_{2-x}$Na$_x$CuO$_2$Cl$_2$ and 
Bi$_2$Sr$_2$Dy$_{0.2}$Ca$_{0.8}$Cu$_2$O$_{8+\delta}$, has revealed intense 
spatial variations in asymmetry of electron tunneling currents with bias 
voltage that forms unidirectional domains coexisting with inhomogeneous 
$d$-wave superconductivity~\cite{Koh07}. In particular, it has been found, 
that the asymmetry occurs primarily at the oxygen sites being indicative 
of a short-range \emph{bond-centered} charge pattern with a period of 
four lattice spacings. 
In this paper we show that the bond-centered modulation observed in the STM 
experiments might be naturally interpreted in terms of a valence bond crystal, 
i.e., spin-rotationally invariant phase with spatially varying bond charge 
hopping and a concomitant modulation of short-range AF 
correlations~\cite{Rac07a}. 

\section{Renormalized mean-field theory}
We begin by discussing RMFT of the $t$-$J$ model applied to the case with
homogeneous charge distribution. In this approach, the Gutzwiller projection 
removing double occupancy is handled with statistical weight factors 
$g_t=2x/(1+x)$ and $g_J=4/(1+x)^2$ which account for different probabilities 
of hopping and superexchange processes in the projected and unprojected 
wave functions. Hence the mean-field Hamiltonian reads,  
\begin{align}
\label{eq:H_MF}
              H = &- t\sum_{\langle ij\rangle,\sigma} g_{ij}^t
                  (c^{\dagger}_{i,\sigma}c^{}_{j,\sigma}+h.c.)
                   -\mu\sum_{i,\sigma}n_{i,\sigma}\nonumber \\
                  &-\frac{3}{4} J \sum_{\langle ij\rangle,\sigma}g_{ij}^J
                  [(\chi_{ji}c^{\dagger}_{i,\sigma}c^{}_{j,\sigma}
                  + \Delta_{ji}c^{\dagger}_{i,\sigma}c^\dagger_{j,-\sigma}
                  + h.c.) -|\chi_{ij}|^2  -|\Delta_{ij}|^2],
\end{align}
with the Bogoliubov-de Gennes self-consistency conditions for the bond-
$\chi_{ji}=\langle c^\dagger_{j,\sigma}c^{}_{i,\sigma}\rangle$ and pair-order
$\Delta_{ji}=\langle c^{}_{j,-\sigma}c^{}_{i,\sigma}\rangle
            =\langle c^{}_{i,-\sigma}c^{}_{j,\sigma}\rangle$
parameters in the unprojected state. Hereafter, we shall assume a typical
value $t/J=3$. 

\begin{figure}[t]
\begin{center}
 \resizebox{0.95\linewidth}{!}{
  \includegraphics*{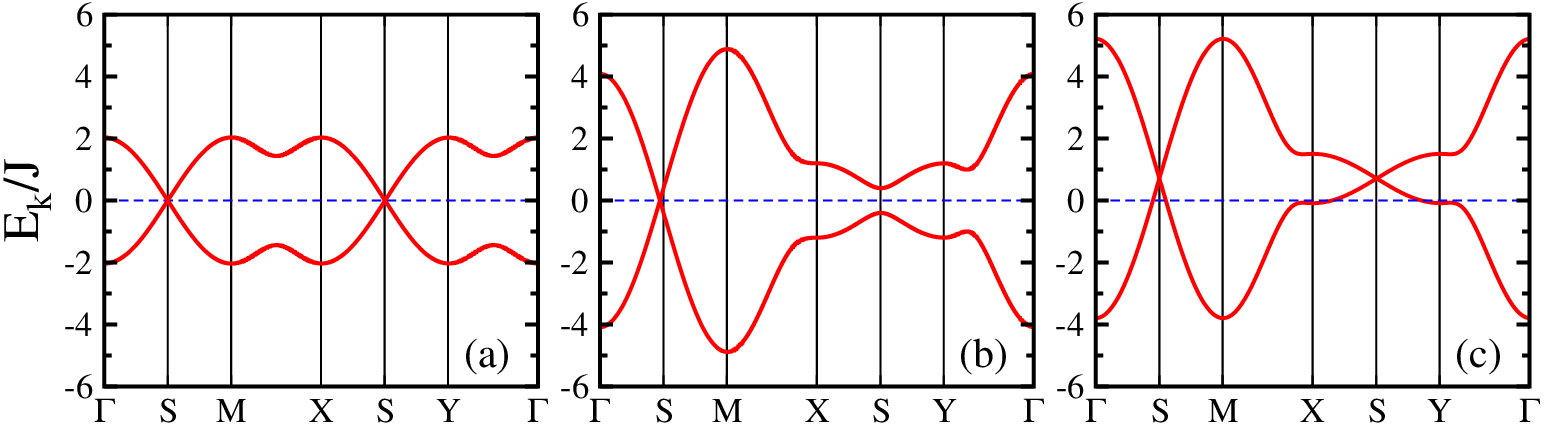}}
\end{center}
 \caption{\label{fig:BS2} 
Electronic structure along the main directions of the Brillouin zone of: 
(a) $d$-wave RVB/SF ($x=0$); 
(b) $d$-wave RVB and (c) SF phase (both for $x=1/8$). }
\end{figure}  

Even though the original proposal for the high-$T_c$ superconductivity 
was the $s$-wave BCS wave function, it immediately turned out that it is 
the $d$-wave BCS state with $\Delta_{ij}=\pm\Delta$ for the nearest-neighbor 
pairs along the $x$ ($y$) axis, respectively, which gives the lowest 
energy~\cite{dwave}. At half-filling, such a phase is equivalent 
to the staggered flux (SF) state with complex 
$\chi_{ij}=|\chi|\exp[(-1)^{i_x+j_y}i\phi]$ yielding circulating
currents whose chirality alternates from plaquette to plaquette~\cite{sfp}. 
In this limit, the $t$-$J$ model reduces to the Heisenberg Hamiltonian
with the local SU(2) gauge symmetry  corresponding to the following 
particle-hole transformation, 
\begin{equation}
 \begin{pmatrix}
  c_{i\uparrow}^{\dagger}    \\
  c_{i\downarrow} 
 \end{pmatrix}
=
 \begin{pmatrix}
  \alpha_i   & \beta_i    \\
 -\beta_i^*  & \alpha_i^* 
 \end{pmatrix}
 \begin{pmatrix}
  c_{i\uparrow}^{\dagger}    \\
  c_{i\downarrow}  
 \end{pmatrix},
\label{eq:H_kin}
\end{equation}
with $\alpha_i\alpha_i^{*}+\beta_i\beta_i^{*}=1$. It mixes an $\uparrow$-spin 
particle with a $\downarrow$-spin hole and hence decouplings in terms of 
$\Delta$ or $\chi$ become indeed equivalent. 
In order to appreciate this better let us consider the related
Hamiltonian matrices using the Bogoliubov-Nambu formalism with  
$\eta_{\bf k}=(c^{}_{{\bf k}\uparrow},c^{\dag}_{-{\bf k}\downarrow})$  
for the $d$-wave RVB phase and 
$\eta_{{\bf k}}=(c^{}_{{\bf k},\sigma},c^{}_{{\bf k}+{\bf Q},\sigma})$ 
with ${\bf Q}=(\pi,\pi)$ for the SF phase: 
\begin{equation}
M_{\bf k}^{\rm RVB}=
 \begin{pmatrix}
  -\varepsilon_{\bf k} -\mu   & \Delta_{\bf k}    \\
   \Delta_{\bf k}             & \varepsilon_{\bf k} +\mu   
 \end{pmatrix},
\qquad
M_{{\bf k}}^{\rm SF}=
 \begin{pmatrix}
  -\varepsilon_{\bf k} -\mu   & i\chi_{\bf k}    \\
  -i\chi_{\bf k}             & \varepsilon_{\bf k} -\mu   
 \end{pmatrix},
\label{eq:H}
\end{equation}
where $\varepsilon_{\bf k}=(tg^t+\tfrac{3}{4}Jg^JRe\chi)\gamma_{+}$, 
$\Delta_{\bf k}=\tfrac{3}{4}Jg^J\Delta\gamma_{-}$, 
$\chi_{\bf k}=\tfrac{3}{4}Jg^JIm\chi\gamma_{-}$ with 
$\gamma_{\pm}=2(\cos k_x\pm\cos k_y)$. Therefore, the corresponding spectra 
are given by:
\begin{equation}
E_{\bf k}^{RVB}=   \pm\sqrt{(\varepsilon_{\bf k}-\mu)^2+\Delta_{\bf k}^2}, 
\qquad {\rm and} \qquad  
E_{\bf k}^{SF}=-\mu\pm\sqrt{\varepsilon_{\bf k}^2+\chi_{\bf k}^2}. 
\end{equation}
At half-filling ($\mu=0$), one finds $\Delta=\chi=0.169$  ($|\chi|=0.239$ and 
$\phi=\pi/4$) for the $d$-wave RVB (SF) phase, respectively. Hence, in the
latter case $Re\chi=Im\chi=0.169$ and, since $\Delta_{\bf k}=\chi_{\bf k}$, 
both spectra become degenerate. As shown in Fig.~\ref{fig:BS2}(a), the key 
feature of the obtained spectrum is that the energy gap vanishes linearly along 
the $S=(\pi/2,\pi/2)$ point forming a cone-like dispersion. 
While this cone remains pinned to the Fermi surface in the $d$-wave RVB phase, 
finite doping takes the node of the SF order away from the Fermi surface 
and opens hole pockets around the $S$ point [see Fig.~\ref{fig:BS2}(b,c)].
Nevertheless, both excitation spectra remain similar, which makes 
the SF phase an excellent candidate for the normal pseudogap phase that
emerges below a characteristic temperature $T^*$. Moreover, short-range 
staggered orbital current-current correlations have been found in the 
Gutzwiller-projected $d$-wave RVB phase~\cite{Iva00}, in the exact ground 
state of the $t$-$J$ model with a negative two-hole binding
energy~\cite{Leu00}, as well as by analyzing motion of a hole pair in the 
AF background~\cite{Wro01}. 

\begin{figure}[t]
\begin{center}
 \resizebox{0.7\linewidth}{!}{
  \includegraphics*{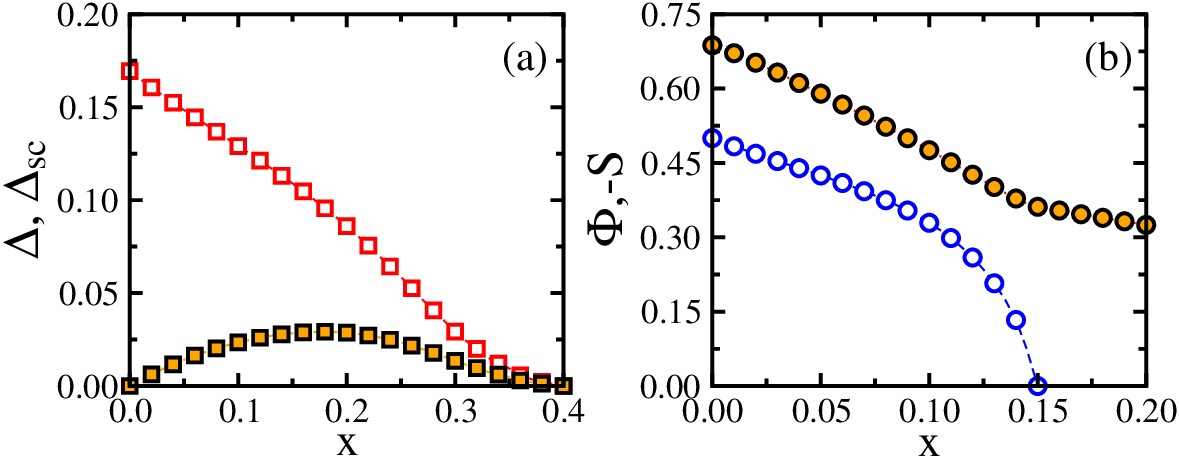}}
\end{center}
 \caption{\label{fig:BS} Doping dependence of: 
(a) pairing amplitude $\Delta$ (open squares) and SC order parameter 
    $\Delta_{\rm SC}$ (solid squares) in the $d$-wave RVB phase, 
    as well as
(b) plaquette flux $\Phi$ (open circles) and spin correlations ${\mathcal S}$ 
    (solid circles) in the SF phase. }
\end{figure}

Next, as shown in Fig.~\ref{fig:BS}(a), variational parameter $\Delta$ 
is the largest at $x=0$ and vanishes linearly with doping. In contrast, 
$g^t$ is an increasing function of $x$ so that the resulting SC order 
parameter $\Delta_{\rm SC}=g^t\Delta$ reproduces qualitatively the SC dome. 
Finally, Fig.~\ref{fig:BS}(b) depicts the doping dependence of the   
fictitious flux (in unit of the flux quantum) defined by a sum over the 
four bonds of the plaquette 
$\Phi_{\Box}=\frac{1}{2\pi}\sum_{\langle ij\rangle\in\Box}\Theta_{ij}$ 
as well as the AF spin correlations ${\mathcal S}=-\frac{3}{2}g_{}^{J}|\chi|^2$. 
Here, the appearance of a finite flux at $x\simeq 0.15$ clearly strengthens 
${\mathcal S}$ with respect to the Fermi liquid state where 
$\chi$ is entirely real. 

\section{Unidirectional charge order}

We turn now to the discussion of bond-centered (with a maximum of the hole 
density spread over two-leg ladders) inhomogeneous RVB (chiral) states 
derived from the parent $d$-wave RVB (SF) phases, respectively.
Hereafter we refer to the former as $\pi$-phase domain RVB phase
($\pi$DRVB), as it involves two out-of-phase SC domains 
(see also Ref.~\cite{Ogata}), separated by horizontal bonds with vanishing 
pairing amplitudes, named as ``domain wall'' (DW), where $\Delta_{ij}$ gains 
a phase shift of $\pi$. Similarly, due to the existence of DWs which act 
as nodes for the staggered current and introduce into the SF order parameter 
a phase shift of $\pi$, we refer to the latter as $\pi$DSF state. 

We consider both original ($q=0$) and modified ($q=1$) Gutzwiller factors 
depending on local hole densities $n_{hi}$,
\begin{align}
  g_{ij}^J &=\frac{4(1-n_{hi})(1-n_{hj})}
                   {\alpha_{ij}+q[8n_{hi}n_{hj}\beta_{ij}^{-}(2)
                   +16\beta_{ij}^{+}(4)]},
\label{eq:GJ} \\
  g_{ij}^t &=\sqrt{\frac{4n_{hi}n_{hj}(1-n_{hi})(1-n_{hj})}
                   {\alpha_{ij}+q[8(1-n_{hi}n_{hj})|\chi_{ij}|^2
                   +16|\chi_{ij}|^4]}},
\label{eq:Gt}
\end{align}
with $\alpha_{ij}=(1-n_{hi}^2)(1-n_{hj}^2)$ and
$\beta_{ij}^{\pm}(n)= |\Delta_{ij}|^n\pm|\chi_{ij}|^n$. Note, however, 
that the Gutzwiller renormalization scheme becomes substantially more 
complicated in the case of inhomogeneous charge distribution as the local 
density may change before/after projection~\cite{Gro07}. As a consequence, 
Eqs. (\ref{eq:GJ}) and (\ref{eq:Gt}) may provide only an approximate way 
of the Gutzwiller projection. Finally, using the unit cell translation 
symmetry~\cite{Rac06}, calculations were carried out on a large $256\times
256$ cluster at a low temperature $\beta J=500$.

\begin{figure}[t]
\begin{center}
 \resizebox{0.95\linewidth}{!}{
  \includegraphics*{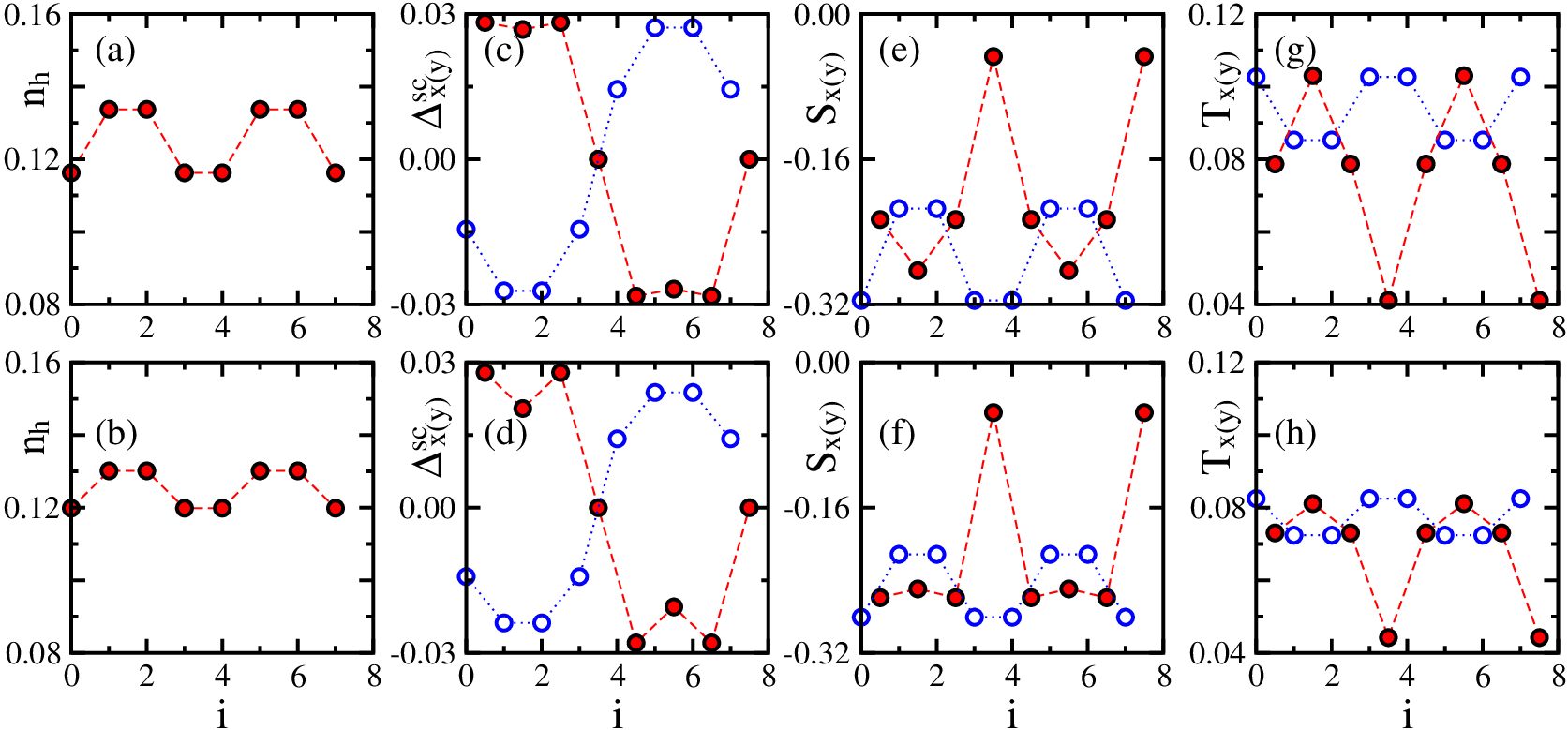}}
\end{center}
 \caption{\label{fig:profRVB} 
(a,b) Hole density $n_{hi}$,
(c,d) SC order parameter $\Delta_{i\alpha}^{\rm SC}$,
(e,f) spin correlation ${\mathcal S}_{i}^{x}$, and
(g,h) bond charge ${\mathcal T}_{i}^{y}$, found in the $\pi$DRVB phase.
Top (bottom) panels depict the results obtained using original (modified) 
Gutzwiller factors;
solid (open) circles in panels (c-h) correspond to the $x$ ($y$) 
direction, respectively.
}
\end{figure}

\begin{figure}[t]
\begin{center}
 \resizebox{0.95\linewidth}{!}{
  \includegraphics*{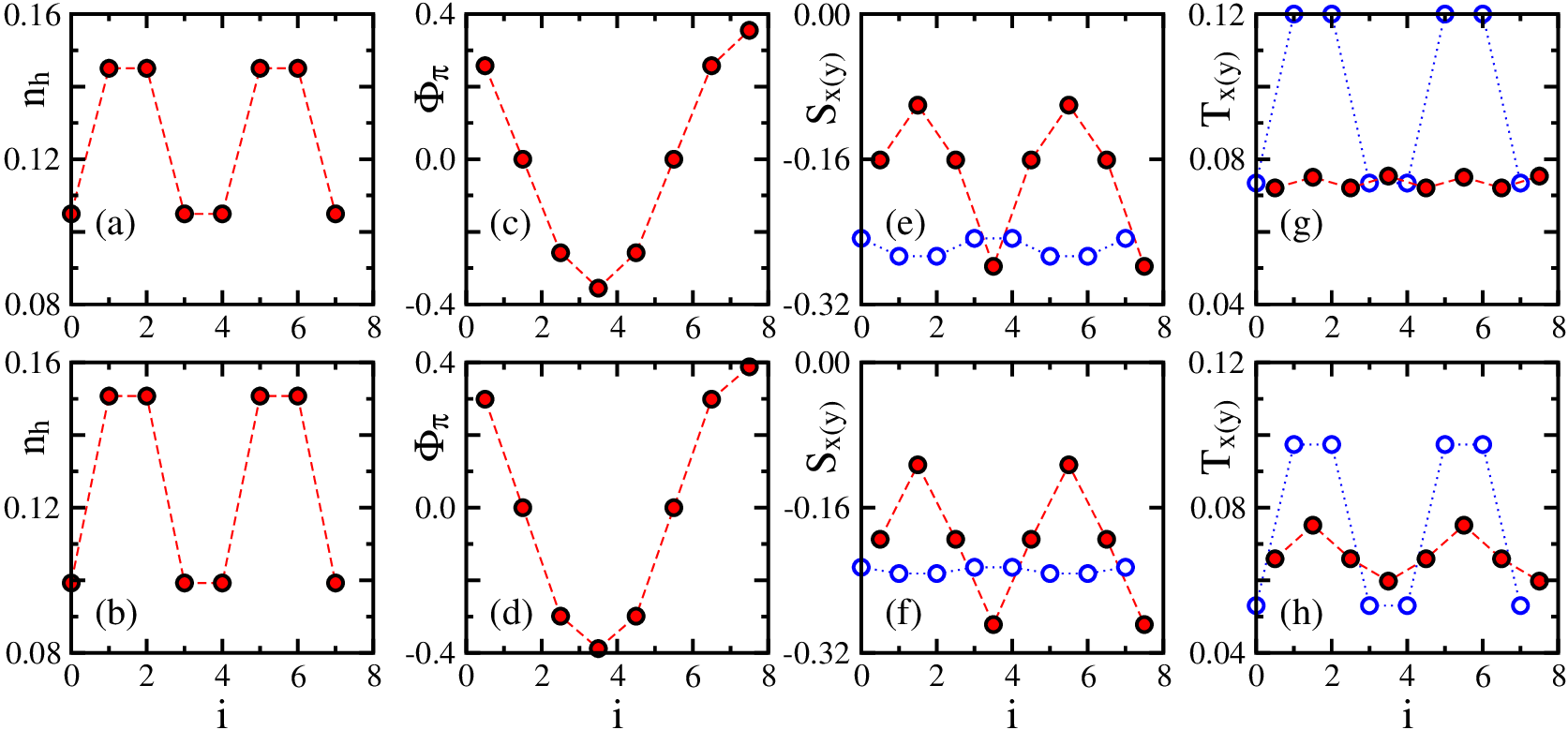}}
\end{center}
 \caption{\label{fig:profSF} 
(a,b) Hole density $n_{hi}$,
(c,d) modulated flux $\Phi_{\pi i}$,
(e,f) spin correlation ${\mathcal S}_{i}^{x}$, and
(g,h) bond charge ${\mathcal T}_{i}^{y}$, found in the $\pi$DSF phase.
Top (bottom) panels depict the results obtained using original (modified) 
Gutzwiller factors; solid (open) circles in panels (e-h) correspond to 
the $x$ ($y$) direction, respectively.
}
\end{figure}

The corresponding stripe profiles in both phases  shown in 
Figs.~\ref{fig:profRVB} and \ref{fig:profSF} are clearly a compromise between 
the superexchange energy $E_J$ and kinetic energy $E_t$ of doped holes. 
On the one hand, a reduction of the SC or flux order parameters
(the latter known to frustrate coherent hole motion~\cite{flux}) enables 
a large bond charge hopping 
${\mathcal T}_i^{y}=2g_{i,i+y}^{t}Re\{\chi_{i,i+y}\}$ along the DWs 
as in the usual stripe scenario~\cite{Rac06}. On the other hand, 
it simultaneously results in the suppression of the AF correlations 
${\mathcal S}_i^{x}=
-\frac{3}{2}g_{i,i+x}^{J}(|\chi_{i,i+x}|^2+|\Delta_{i,i+x}|^2)$ 
along the transverse bonds.

However, a closer inspection of Figs.~\ref{fig:profRVB} and \ref{fig:profSF} 
as well as Table~I indicates that this competition is especially 
subtle in the $\pi$DRVB state. Indeed, instead of increasing hole level 
in the SC areas in order to reinforce the SC order parameter, 
the system prefers a more spread out charge distribution, which suggests that 
the $d$-wave RVB state is less disposed to phase separation than the SF one
where in fact also other charge instabilities have been found~\cite{dp,Rac07}. 
Moreover, as listed in Table~I, both $E_t$ and $E_J$ are reduced with respect 
to the uniform $d$-wave RVB superconductor. In contrast, $\pi$DSF phase fully
optimizes both energy contributions simply by expelling holes from the regions 
between the stripes and accommodating them at the DWs. Indeed, a low local
doping level strengthens plaquette flux which reaches the value 
$\Phi_{\Box}\simeq 0.35$ expected for $x\simeq 0.1$ [see Fig.~\ref{fig:BS}(b)]. 
Moreover, the holes accommodated at the DWs enhance locally Gutzwiller 
factors $g_{ij}^{t}$ and allow the phase to retain a favorable $E_t$. 
Taken together, these two effects are responsible for a much stronger charge 
modulation of the $\pi$DSF phase as compared to its $\pi$DRVB counterpart.  

\begin{table}[t]
{\small
\rightline{TABLE I}
\noindent
RMFT kinetic energy $E_t$, magnetic energy $E_J$, and free energy $F$ of
the locally stable phases: $\pi$DSF, SF, $\pi$DRVB, and $d$-wave RVB one 
at $x=1/8$.
\begin{center}
\begin{tabular}{c|c|c|c|c|c|c}
\hline\hline
\multicolumn{1}{c}{}                      &
\multicolumn{3}{c|}{ original $g_{ij}$ }  &
\multicolumn{3}{c}{ modified $g_{ij}$ }   \\
\hline
 phase        &      $E_t/J$   &    $E_J/J$    &    $F/J$  
              &      $E_t/J$   &    $E_J/J$    &    $F/J$  \\
\hline
 $\pi$DSF     &      $-$1.0252&    $-$0.4320  &  $-$1.4572
              &      $-$0.8514&    $-$0.4269  &  $-$1.2783 \\
 SF           &      $-$1.0345&    $-$0.4246  &  $-$1.4591        
              &      $-$0.8622&    $-$0.4230  &  $-$1.2852 \\
 $\pi$DRVB    &      $-$1.0160&    $-$0.4607  &  $-$1.4767
              &      $-$0.8719&    $-$0.4518  &  $-$1.3237 \\
 RVB          &      $-$1.0232&    $-$0.4838  &  $-$1.5070
              &      $-$0.8863&    $-$0.4784  &  $-$1.3647 
\end{tabular}
\end{center}
}
\end{table}

Unfortunately, the total RMFT energy in both phases differs substantially from 
the one obtained within the VMC scheme: $E_{\rm DRVB}/J\simeq -1.34$ 
($E_{\rm DSF}/J\simeq-1.33$) for the  $\pi$DRVB ($\pi$DSF) phase, 
respectively~\cite{Rac07a}. 
We consider therefore the so-called modified Gutzwiller factors 
where the effects of the nearest-neighbor correlations $\chi_{ij}$ and 
$\Delta_{ij}$ are also included~\cite{RMF}. 
First of all, one observes that the inclusion of the intersite correlations 
weakens (strengthens) stripe order in the $\pi$DRVB ($\pi$DSF) phase, 
respectively. Indeed, in both cases the holes are ejected from the regions 
in between stripes into the DWs defined as nodes of the SC/flux order 
parameter. Similar tendency towards accommodating the holes at the DWs 
has also been established in the VMC calculations~\cite{Man08}. Furthermore, 
while the short-range AF correlations remain either unaltered or they are 
changed in such a way that $E_J$ remains almost constant, 
modified Gutzwiller factors mainly renormalize bond charge hopping.   
As a consequence, the total energy in both phases approaches the one found 
in the VMC scheme (see Table~I).

In summary, we believe that our results provide insights into the formation 
of the recently found bond-centered charge order that coexists with modulated 
$d$-wave superconductivity. Moreover, we expect a further enhancement 
of the proposed valence bond crystal near impurities that break the space 
group symmetry of the $t$-$J$ Hamiltonian by producing a modulation in the 
magnitude of the superexchange coupling~\cite{Sach07}. In fact, it has 
recently been argued that the dopant-induced spatial variation of the atomic 
levels indeed strengthens locally the AF superexchange interaction~\cite{Mas07}.

\section*{Acknowledgments}
The author acknowledges support from the Foundation for Polish Science (FNP), 
Minist\`ere Fran\c{c}ais des Affaires Etrang\`eres under Bourse de Recherche, 
as well as from Polish Ministry of Science and Education under Project 
No. N202 068 32/1481.

\end{document}